# Charge migration in Pb(Zr,Ti)O$_3$ ceramics and its relation to ageing, hardening and softening

M.I.Morozov[1] and D.Damjanovic[2,a]

[1]*Institute for Ceramics in Mechanical Engineering, Universtität Karlsruhe (TH), 776131 Karlsruhe, Germany*

[2] *Ceramics Laboratory, Swiss Federal Institute of Technology - EPFL, 1015 Lausanne, Switzerland*

*Abstract* — The dielectric response of hard (Fe-doped) and soft (Nb-doped) rhombohedral Pb(Zr$_{0.58}$Ti$_{0.42}$)$_{1-x}$Me$_x$O$_3$ (PZT; Me= Fe or Nb) ceramics was studied at subswitching conditions over a wide range of temperatures (50°C to 450 °C) and frequencies (10 mHz to 10 kHz). The results show qualitative differences in behavior of the acceptor and donor doped samples. Hard materials exhibit a steep increase of the complex permittivity with decreasing frequency. The onset of the dispersion is thermally activated with activation energy of about 0.6-0.8 eV and is attributed here to oxygen vacancy hopping. Activation energy for ac conductivity observed in soft materials is estimated to about 1.7 eV, corresponding to the half of the energy gap of Pb(Zr,Ti)O$_3$ and is thus consistent with electronic conduction. The relevance of ionic hopping conductivity in hard materials to ferroelectric aging / deaging and hardening is analyzed. Strong ionic conductivity in hard and its absence in soft samples agree well with the dipolar mechanism of ageing in hard materials and the absence of significant ageing in soft materials.
PACS: 77.22.Gm 77.84.Dy 72.80.Sk 72.20.Ee

---

[a] Electronic mail: dragan.damjanovic@epfl.ch

# 1. Introduction

Hardening and softening of ferroelectric materials by doping are perhaps the most important tools to control their electro-mechanical properties. Perovskite ferroelectrics can be soften if material is doped with donor-type dopants and harden by the acceptor-type dopants.[1] Under usual operating conditions soft materials exhibit large electromechanical coefficients, square hysteresis, large nonlinearity, weak aging, low conductivity and high dielectric losses while the hard ferroelectrics exhibit strong aging, pinched hysteresis loop in aged unpoled state, weak nonlinearity, lower coupling coefficients, low dielectric losses and moderate conductivity. Whereas softening of the ferroelectric properties is still not clearly understood,[2] the hardening is intimately related to aging, i.e. gradual hardening of ferroelectric properties can be evidenced with time.[3] Understanding of hardening as a result of aging is nowadays much advanced in comparison with the softening phenomenon in spite of the fact that several contradicting theories of aging exist.[4-8]

The aging (evolution of the properties with time upon changing sample's thermal, electrical or mechanical environment) originates principally from stabilization of domain structure and immobilization (clamping) of domain walls.[9] With respect to interaction between mobile charge carriers and the spontaneous polarization, the following three mechanisms of the domain wall immobilization in hard ferroelectrics are the most often discussed in the literature.[10] The first is the volume (or bulk) effect, which assumes alignment (or ordering) of microdipoles formed by charged point defects with ferroelectric polarization in the material bulk between adjacent domain walls (e.g., $Fe'_{Ti} - V_O^{\bullet\bullet}$; we use Kröger-Vink notation for defects[11]).[12-23] The dipoles reorientation takes place through displacement of oxygen vacancies.[4,20,22,23] Even though minimization of the electrostatic energy is usually discussed in the literature as the only driving force for the defect orientation, the minimization of the elastic energy associated with coupling of the elastic defect "dipoles" and lattice deformation can be significant.[24-26] The volume effect has recently been studied in detail by Ren and coworkers[4,27] and has been referred to as "symmetry conforming" mechanism. Displacement of $V_O$ within oxygen octahedron as a contributing factor to ageing is also supported by *ab initio* calculations.[28] The second mechanism of domain walls immobilization is known as the domain wall effect, which assumes diffusion of defects (such as electrons or oxygen vacancies) into charged domain walls, thus fixing their position.[29-33] The third mechanism is related to the interface effects, where space charges associated with defects collect at the grain boundaries, at the external perimeter of the crystal or other interfaces.[6-8,34]

The common feature of all abovementioned processes is the electrostatic rearrangement of charged carriers and their interaction with polarization, while the difference is in the final location of mobile charges. Thus, proposed scenarios of aging and deaging assume different distances for charge displacement: over a unit cell (bulk effect), over a domain width (domain wall effect), or over a part of grain or crystal (interface effect). More recently, Li et al.[35] suggested that mechanisms for domain wall structure stabilization in hard PZT and new, lead free $(K,Na,Li)NbO_3$ ceramics are different, the bulk effect being dominant in PZT and long-range migration of charges in the lead free material. It has been hinted in a previous study by the authors[36,37] that more than one process of ageing may be active in a material: in hard PZT the bulk effect plays a key role, but the long-range charge migration may contribute to the deaging under strong electric fields.

An aged hard ferroelectric can be "deaged" by destroying the electrostatic order. This can be accomplished, for example, by cycling with an electric field[10] quenching from a high temperature,[36,38,39] or by illumination.[31] Deaging of hard PZT by electric field is the best documented case.[10,37] For a fixed amplitude of the applied field the deaging time depends on temperature and cycling field frequency, and can be presented as an Arrhenius-type process with characteristic activation energy.

Jaffe et al.[1] have interpreted the absence of ageing in soft materials by the increased mobility of domain walls with respect to undoped PZT. Higher domain walls mobility was suggested to aid quick relief of internal stress upon removal of the poling field.[40] If this is indeed so, it would be more correct to speak of rapid ageing in soft materials rather than its absence. As over longer times the two mean practically the same thing, we shall use here term "absence of aging" to contrast better hard and soft materials. If softening is a result of the higher domain wall mobility, the absence of ageing and softening seem to be intimately connected, just like ageing and hardening. Why "soft" materials have higher domain wall mobility than undoped materials of the same composition is presently not clear and several hypotheses have been advanced without theoretical or experimental proofs, including the following: (i) donor dopants compensate effects of acceptor cations that are naturally present in undoped materials,[1] (ii) lead vacancies assumed to compensate donor dopants (e.g., $Nb^{\bullet}_{Ti} - V''_{Pb}$) help to reduce internal stresses in ceramics and make domain walls more mobile,[1,40] and (iii) softening is related to electron transfer between defects thus minimizing the space charges at domain walls.[41] Clearly, charge transport (or its absence) is expected to be an important indicator of ageing, hardening and softening processes.

In this paper we present results of an experimental study of charge migration processes in hard, undoped and soft PZT ceramics. The charge migration was investigated using dielectric spectroscopy at temperatures from 50° to 450°C and frequency range from $10^{-2}$ to $10^4$ Hz. As other methods used to study dielectric response, the one employed here measures the total charge displacement and cannot distinguish directly between contributions from the ac conductivity (e.g., charge hopping) and polarization mechanisms (e.g., domain wall displacement). The used method, however, allows distinguishing between the ac processes associated with a short-distance migration of charge carriers and the quasi dc processes associated with migration of charge carriers over longer distances. In hard materials, both the ac and quasi-dc processes reveal themselves as thermally-activated phenomena whose activation energies can be compared with those for deaging process.[10,37] Thus, we compare here the activation of the ac migration processes in hard PZT obtained in experiments performed at subswitching fields and in the frequency domain, with the activation of deaging in the same type of ceramics at switching conditions in the time domain using data from Ref. 10. In agreement with other studies[12,15,19,23] our analysis of charge migration and ageing / deaging suggests that the defect dipole scenario (the bulk effect) is the most likely origin of the "pinched loop" – perhaps the best known manifestation of ferroelectric hardening in Fe-doped PZT ceramics. However, our earlier study of nonlinear dielectric response[37] gives evidence in agreement with recent theoretical results[6] that the hardening (ageing) and deaging processes may be assisted by a longer-range displacement of defects. The link between the conductivity and ageing suggests dominant ionic conduction in hard PZT materials. The present data further demonstrate that the conduction in soft ceramics is low below Curie temperature ($T_C$) and that electron conduction may dominate the conductivity around and above $T_C$. Low conductivity below the $T_C$ is in agreement with the absence of significant aging in soft materials.

**2. Materials preparation**

Undoped, $Fe^{+3}$–doped (hard), and $Nb^{+5}$–doped (soft) rhombohedral $Pb(Zr_{0.58}Ti_{0.42})O_3$ ceramics were prepared by conventional solid state process using standard mixed oxide route. Processing details can be found in Ref. 36. A composition away from the morphotropic boundary was chosen to avoid complications with presence of mixed or monoclinic phases, and because the domain wall structure is simpler than in morphotropic phase boundary region. The base composition was chosen the same as in Ref. 10 in order to use results from this source for comparison and analyses. The dopant substitution was assumed to be on (Zr,Ti) site so that the nominal formula of the samples is $Pb(Zr_{0.58}Ti_{0.42})_{1-x}Me_xO_3$ where Me is either Fe or Nb and x = 0.1 (only Fe), 0.2 (only Nb), 0.5 and 1.0 at % (abbreviated 58/42 PZT with x at% Me). The ionic charge compensating defects[1] (lead vacancies $V''_{Pb}$ for $Nb^{\bullet}_{Ti}$ and oxygen vacancies $V^{\bullet\bullet}_O$ for $Fe'_{Ti}$) are assumed to form spontaneously during thermal treatments (calcination and sintering). This is possible in PZT because of high volatility of PbO.[1] It should be noted that thus formed dipoles ($Nb^{\bullet}_{Ti} - V''_{Pb}$ and $Fe'_{Ti} - V^{\bullet\bullet}_O$) are not neutral and that it is not clear which charged species compensate for charge unballance.[42] This point illustrates well complexity of PZT. Presence of $Fe'_{Ti} - V^{\bullet\bullet}_O$ dipoles in hard material is well established, while defect structure of donor doped PZT is much less understood.

Stoichiometric quantities of powders (adjusted for predetermined weight loss during heating) were calcined in lead oxide saturated alumina crucibles covered by alumina plates. Powders were milled and sieved before and after calcinations. The sintering was performed on pellets uniaxially pressed at 40 MPa and packed into covered alumina crucibles whose volume was a little larger than the volume of pellets. The inner space of the crucibles was filled up with the powder of the same composition as the pressed pellet in order to prevent intensive lead oxide evaporation during the thermal treatment. Ageing of samples was assured by their slow cooling within the furnace down to room temperature, followed by several days of ageing at room temperature. Gold electrodes were deposited on major faces of disk-shaped samples by sputtering.

## 3. Dielectric spectroscopy

Among several methods commonly employed to study dielectric properties in time or frequency domain[43] we use in this study measurements of ac charge generated by applying on sample a periodic voltage signal with fixed frequency, $V = V_0 \sin(\omega t)$. The measurement is repeated at different frequencies in the range from $10^{-2}$ to $10^4$ Hz and temperatures ranging from 50 °C to 450°C for soft and to 250°C for hard ceramics. The Curie temperature of these compositions is around 360°C. For the sake of completeness and to facilitate the reading we repeat here basic relations of dielectric spectroscopy.[44,45]

The total measured current density $J = I/A$, where $A$ is the sample surface area and $I$ the total measured current, consists of two contributions. The first is movement of free charges $\sigma_0 E$, where $\sigma_0$ is the dc conductivity, and $E = V/d$ is the electric field across the sample with thickness $d$. The second contribution is the dielectric displacement current $\partial D/\partial t$, associated with the polarization response $D = \varepsilon_0 E + P$. Thus, $J = \sigma_0 E + \partial D/\partial t$. The former, purely resistive component, is in phase with voltage while the latter, capacitive component, is out of phase with voltage by 90°[44,46]. Following the established practice, the total current can be written in complex form as:[44,45]

$$J = i\omega\varepsilon_0\tilde{\varepsilon}(\omega)E(\omega) = i\omega\varepsilon_0\{\varepsilon'(\omega) - i[\varepsilon''(\omega) + \sigma_0/(\varepsilon_0\omega)]\}E(\omega) \qquad (1)$$

where $E(\omega) = E_0\exp(i\omega t)$, $\varepsilon_0$ is the permitivity of vacuum, $\tilde{\varepsilon}(\omega)$ is the effective (measured) relative dielectric permittivity and $\varepsilon = \varepsilon' - i\varepsilon''$ is the complex relative dielectric permittivity. For simplicity we shall interchange terms permittivity and relative permittivity. The effective permittivity is introduced since the measuring instruments cannot discriminate between the true dielectric response, which does not contain $\sigma_0$, and the effective (measured) response that does. Note also that besides the "true" polarization mechanisms, $\varepsilon$ includes contributions from any ac motion of charges, such as hopping conductivity.[44,45] Equation (1) can be further modified by subtracting, if known, the "instantaneous" dielectric response, $\varepsilon_\infty$ which is implicitly contained in $\tilde{\varepsilon}(\omega)$.[45]

The ac conductivity is defined by:

$$\sigma(\omega) = i\omega\varepsilon_0\tilde{\varepsilon}(\omega) = \omega\varepsilon_0\varepsilon''(\omega) + \sigma_0 + i\omega\varepsilon_0\varepsilon'(\omega) \qquad (2)$$

One sees that the real part of the measured current (in phase with voltage) contains both the dc conductivity and the dielectric loss (for example, from domain wall motion) while the imaginary component (in quadrature with voltage) contains only the dielectric response (including all charge displacement mechanism that contribute to polarization, such as charge-carriers hopping). Separation between the conductivity and the dielectric response is therefore not straightforward. In complex materials, such as ferroelectric ceramics, the direct conductivity may contribute to both $\varepsilon'$ and $\varepsilon''$ through Maxwell-Wagner effects,[46] while charge hopping can manifest itself as dipolar dielectric response.[44] We shall return to this point in more detail in Section 4.

To determine the effective permittivity a periodic low voltage signal generating the electric field with amplitude $E_0$ was applied to electroded samples and the resulting charge response $Q$ was measured using a charge amplifier (Kistler 5011B). $E_0$ was on the order of 0.03 kV/cm. The output signal of the charge amplifier was analyzed with a lock-in amplifier (Stanford Research SR830), which also served as the voltage source. $\tilde{\varepsilon}(\omega)$ was calculated from Eq. (1) and $I = \partial Q(\omega)/\partial t$ as $\varepsilon_0\tilde{\varepsilon}(\omega) = Q(\omega)/(E_0 A)$. With $Q = Q_0(\omega)\exp[i(\omega t - \delta)]$, where $\delta$ is the frequency dependent phase angle between the charge and field, it follows that $\varepsilon_0 \tilde{\varepsilon}'(\omega) = Q_0(\omega)\cos\delta(\omega)/(E_0 A)$ and $\varepsilon_0 \tilde{\varepsilon}''(\omega) = Q_0(\omega)\sin\delta(\omega)/(E_0 A)$. During the measurements samples were located in a small-furnace. The temperature of the sample was measured by a thermocouple placed in its vicinity.

In general, the dielectric response of ferroelectrics is nonlinear and, therefore, contains higher harmonics. All experimental results presented in this article are measured only at the first harmonic, which is rather a standard practice when nonlinear effects are assumed to be small.[43] Nonlinear effects in these materials are reported elsewhere. [37,38]

## 4. Results and discussion

As an illustration of typical dielectric response observed in this work, the frequency dispersion of $\tilde{\varepsilon}(\omega)$ for a hard (1%Fe) and a soft (1% Nb) ceramic is shown in Figs.1a and 1b. The same data are presented as ac conductivity [see Eq. (2)] later in this section. Qualitatively, our results are similar to those reported for compositionally inhomogeneous[47] and compositionally graded PZT ceramics.[48] In those studies the ac conductivity (measured from 10 Hz to 2 MHz in Ref.[48]) and complex permittivity (measured from 1 mHz to 10 kHz, Ref.[47]) are interpreted in terms of "universal" power-law (complex permittivity follows approximately $\omega^n$).[44] In the case of samples investigated here, the "universal" law does not, in general, describe the data well quantitatively. The dispersion observed in our work is qualitatively different from the Debye-like dispersion reported by Verdier et al.[49] in electric-field fatigued PZT ceramics.

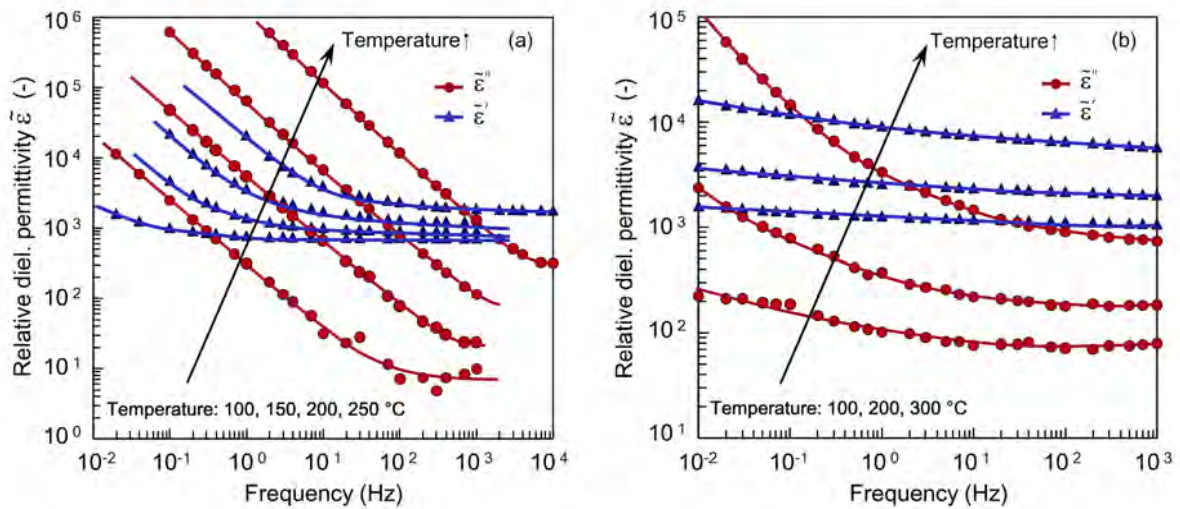

Figure 1 (color online). The real (triangles) and imaginary (circles) parts of the relative effective dielectric permittivity versus driving field frequency measured at various temperatures for (a) hard (1 at% Fe-doped) and (b) soft (1 at% Nb-doped) PZT 58/42 ceramics.

When dielectric properties are measured over a limited frequency range (here, four to five orders of magnitude) there are inherent difficulties in identification of different contributions to the permittivity from either $\tilde{\varepsilon}(\omega)$ or $\sigma(\omega)$ plots. This problem was discussed in detail in Ref. 44 and main arguments are outlined below.

Rapid increase of complex $\tilde{\varepsilon}(\omega)$ with decreasing frequency (as seen in Fig. 1a) can be explained by several different mechanisms, including the following. The first possibility is when dc conductivity and dielectric relaxation are unrelated. In that case [see eq. (1)] the conductivity contributes only to $\tilde{\varepsilon}''(\omega)$, which diverges as $\omega$ approaches zero while $\tilde{\varepsilon}'(\omega)$ is constant or may be subject to an independent dielectric relaxation mechanisms but reaches a finite value at the dc limit. A peak in loss is unrelated to the conductivity. The second case, the Maxwell-Wagner mechanism,[44,46] is related to the presence in the sample of inhomogeneities with different conductivities. The inhomogeneities can be related to secondary phases, grain boundaries or to the near-electrode layer. Even if the conductivities of each phase do not contribute directly to the polarization (i.e., unlike charge hopping), both $\tilde{\varepsilon}'(\omega)$ and $\tilde{\varepsilon}''(\omega)$ of the sample are functions of the conductivities of constituent phases. As in the first case, $\tilde{\varepsilon}'(\omega) = const.$ and $\tilde{\varepsilon}''(\omega)$ diverges as $\omega$ approaches zero. The peak in the loss is determined by the conductivities and permittivities of the constituent phases.[46] Any independent dielectric dispersion of each phase is superimposed on the Maxwell-Wagner dispersion. The third case is hopping conduction, which is quite common in disordered materials.[44,45] The same mechanism but with different activation energies contributes to the ac and dc conduction.[45] A clear region with frequency independent dc conductivity should be seen as $\omega \to 0$. $\varepsilon''(\omega)$ exhibits a peak in the transition region and is nearly constant at high frequencies. The peak in $\varepsilon''(\omega)$ is usually masked when data are presented as $\tilde{\varepsilon}''(\omega)$, and its deconvolution is nontrivial. Above the frequency of the loss peak the ac conductivity increases with the frequency. The short-range charge migration contributes to both $\tilde{\varepsilon}'(\omega)$ and $\tilde{\varepsilon}''(\omega)$, and this can be approximately described by the empirical "universal" power law.[44,45] Jonscher proposed that the "universal" law is the case by itself and that it can describe dielectric relaxation in many dielectrics. In the fourth case, reported for some materials, complex permittivity steeply rise with decreasing frequency, but no loss peak is observed down to the lowest measured frequencies.[44]

Because of practical limitations, in most experiments only a part of the spectrum at frequencies above the loss peak (if one exists) is observed, where both $\tilde{\varepsilon}'(\omega)$ and $\tilde{\varepsilon}''(\omega)$ increase with decreasing frequency. Since this behavior could indicate any of the four cases discussed above it is not possible to assign straightforwardly one of them without additional data at lower frequencies, although this is often made in the literature.[50] We do not think, however, that assignment of any of these cases to the observed dielectric dispersion is important for the following discussion. Even if the data agreed better with either Jonscher's type universality or charge hopping models (such as random free energy barrier model), these models, by their very nature, cannot give much information on microscopic processes involved; exactly because of their universal nature, these models predict the same behavior in different materials regardless of the underlying microscopic physics.[45] Thus, we proceed reasoning in the following way.

In the case such as ours where at low frequencies complex permittivity diverges and $\tilde{\varepsilon}''(\omega) >> \tilde{\varepsilon}'(\omega)$ (Fig. 1a), it is usually possible to make a quite general statement, without specifying exact physical process,[44] that the long-range migration of charges contributes to both $\tilde{\varepsilon}'(\omega)$ and $\tilde{\varepsilon}''(\omega)$. Furthermore, the partial agreement of the data with the charge hopping behavior[45] at high frequencies suggests that an additional process may contribute to the dielectric dispersion. We argue that the following mechanisms are involved: the first is ionic conductivity in hard and electronic conductivity in soft PZT, both dominant at low frequencies and elevated temperatures. The electronic conductivity in soft materials becomes evident at temperatures higher than those shown in Fig. 1b and is displayed in subsequent figures. The second mechanism is related to domain walls displacement, and is dominant at higher frequencies and lower temperatures. As demonstrated in Fig. 1, the dielectric response of the hard and soft ceramics is qualitatively different. While it is not possible, on these data alone, to identify definitely the contributing process, we propose the following coherent interpretation of the data, based on the discussion above and the results described in literature.

Hard ceramics exhibit a transition from a weakly dispersive, nearly constant dielectric response at higher frequencies[51] to highly dispersive response at lower frequencies, Fig. 1a. The transition between the two regimes is temperature dependent. In soft materials, the permittivity dispersion is weaker than in hard materials at lower frequencies, while at higher frequencies the real part of the permittivity decreases nearly linearly on the linear-logarithmic scale. In both soft and hard materials the high frequency response, which extends to high MHz range[51] not shown here, can be assigned to domain walls contributions. Indeed, the logarithmic decrease of the dielectric permittivity and piezoelectric coefficients has been reported earlier for soft PZT[52-54] and has been assigned in Refs.52 and 53 to domain walls displacement in a random energy landscape. In hard materials domain walls are known to be less mobile than in soft, but they still respond to excitation field possibly by bending[55] in a reversible, nondispersive way up to high MHz range.[51,56-58]

Identification of the low frequency dispersion in hard and soft ceramics is more challenging. Referring to the Jonscher's classification of dielectric responses in solids[44] and the discussion above, the low frequency response in hard ceramics is most likely related to the conductivity ($\tilde{\varepsilon}'' \gg \tilde{\varepsilon}'$) and can possibly be attributed to the charge hopping.[45] The most likely charge carriers in hard samples are oxygen vacancies, which are known to be mobile in perovskite materials.[20,23] We exclude here as a possibility the charge transfer between $Fe^{+2}$ and $Fe^{+3}$ because electron paramagnetic resonance (EPR) studies report as a prevalent defect in hard PZT positively charged $(V_O^{\bullet\bullet} - Fe_{Ti}')^{\bullet}$ pair, i.e. Fe is probably trivalent.[42,59,60]

The response of the soft ceramics appears to be free from a significant hopping charge contribution almost up to the Curie temperature. This result is in agreement with the early data of Gerson[1,61] who reported that dielectric loss of soft PZT is dominated by domain walls contributions and is nearly free from resistive component at not-too-high temperatures. The presence of hopping conduction in the hard and its absence in the soft ceramics in the same frequency and temperature range, suggest the key role of the mobile charge species in the aging process, which is observed only in hard PZT.[1,62] A significant conductivity in soft samples is observed only at temperatures approaching and above the Curie temperature and may be attributed, as shown below, to electronic conduction.

Finally, we comment on the possibility that the large increase of $\tilde{\varepsilon}(\omega)$ in hard samples with decreasing frequency is due to displacement of domain walls. Comparison of the results of this work with the literature data[63] indicates that creep or sliding of domain walls may contribute to the effective low-frequency dielectric permittivity in a similar fashion as hopping transport of charge defects, making their separation difficult. In PZT, domain walls are considered to be more mobile in soft than in hard materials, thus one would expect stronger domain wall contribution at low frequencies in soft than in hard samples. The absence of strong low-frequency dispersion in soft and its presence in hard materials lead us to suggest that the low frequency dispersion in hard materials is dominated by ionic defect hopping rather than domain-walls displacement. Nevertheless, this issue is nontrivial and requires further studies.

Well pronounced transitions from the dielectric response of the ferroelectric domain walls to the nearly conductive response of hopping charges are observed in hard ceramics at frequencies where the slope of $\tilde{\varepsilon}$ changes strongly (Fig. 1a). The frequency of this transition increases with increasing temperature, as is expected for a thermally activated process. We use this transitional frequency plotted versus temperature in the Arrhenius scales to obtain the activation energy $E_a$ for the hopping charge contribution to the dielectric response. Note that the charge hoping is active already at much higher frequencies[45] and that the activation energy estimated here refers to the frequencies at which charge hopping contribution to the effective permittivity begins to dominate over domain wall contributions. The main difficulty of such analysis lies in determining consistently the transitional frequencies for $\tilde{\varepsilon}''(\omega)$, as they are located at the limits of the characterized region. The expansion of the frequency measurement range is hampered by the instrumental limits. Alternatively, one can analyze the frequency dispersion of the absolute effective dielectric permittivity $|\tilde{\varepsilon}(\omega)|$ (Fig. 2a) which exhibits similar behavior as $\tilde{\varepsilon}''(\omega)$, with transitional frequencies located somewhere in between the ones for $\varepsilon'(\omega)$ and $\tilde{\varepsilon}''(\omega)$. In addition, we also analyze the dispersion of $M''(\omega)$, the imaginary part of the dielectric modulus $M(\omega) = 1/\tilde{\varepsilon}(\omega)$, (Fig. 2b), which exhibits a maximum at transitional frequencies. The dielectric modulus is often used in dielectric spectroscopy of conductive materials.[45,64-67] Characterization of dielectric spectra using the dielectric modulus is subjected to considerable controversy.[45,50,64,65] In this work we use it only as a complementary tool for determination of the activation energy in the transitional region, because the frequency of the maximum in $M''(\omega)$ may be easier to determine than the frequency where $\tilde{\varepsilon}''(\omega)$ changes slope.

The transitional frequencies determined using permittivity and modulus representations exhibit typical Arrhenius behavior, as shown in insets in Fig. 2. The close values of the activation energies $E_a$ obtained from $|\tilde{\varepsilon}(\omega)|$ and $M''(\omega)$ show that our approximation is acceptable.

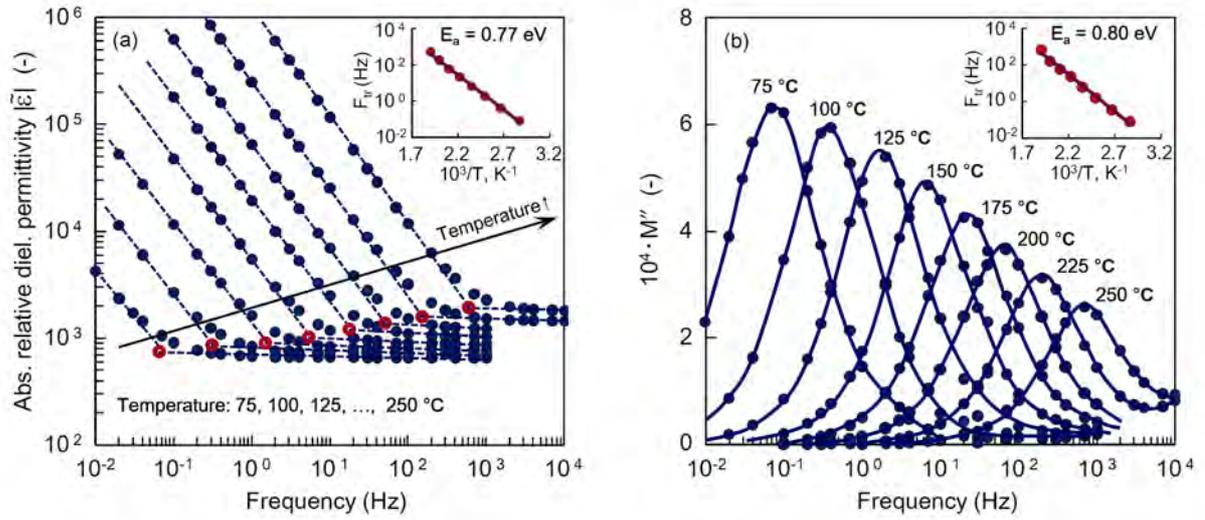

Figure 2 (color online). The frequency dispersion of the absolute value of the complex effective relative permittivity (a) and the dielectric modulus (b) at various temperatures for 1 at% Fe-doped hard PZT (58/42) ceramics. The insets show the temperature dependences of the transition frequencies in the Arrhenius scale and corresponding activation energies. The transition frequencies used for calculation of activation energy in (a) are marked by dots at cross section of the dashed lines.

The analysis of transitional frequencies for $|\tilde{\varepsilon}(\omega)|$ and $M''(\omega)$ is more difficult for the undoped and soft PZT ceramics. In the case of undoped PZT, the $|\tilde{\varepsilon}(\omega)|$ demonstrates a more complex frequency dependence (Fig. 3a) than both Fe- and Nb-doped PZT. It seems that the dispersion involves a multiple transitional processes in which the transition from the nearly constant dielectric response dominated by domain walls displacement at high frequencies (phase angle close to 0°) to the nearly conductive response at lower frequencies (phase angle approaching 90°) is passing through a third dispersion process. Our experimental data in the characterized frequency and temperature ranges are insufficient to interpret the origin of this third process.

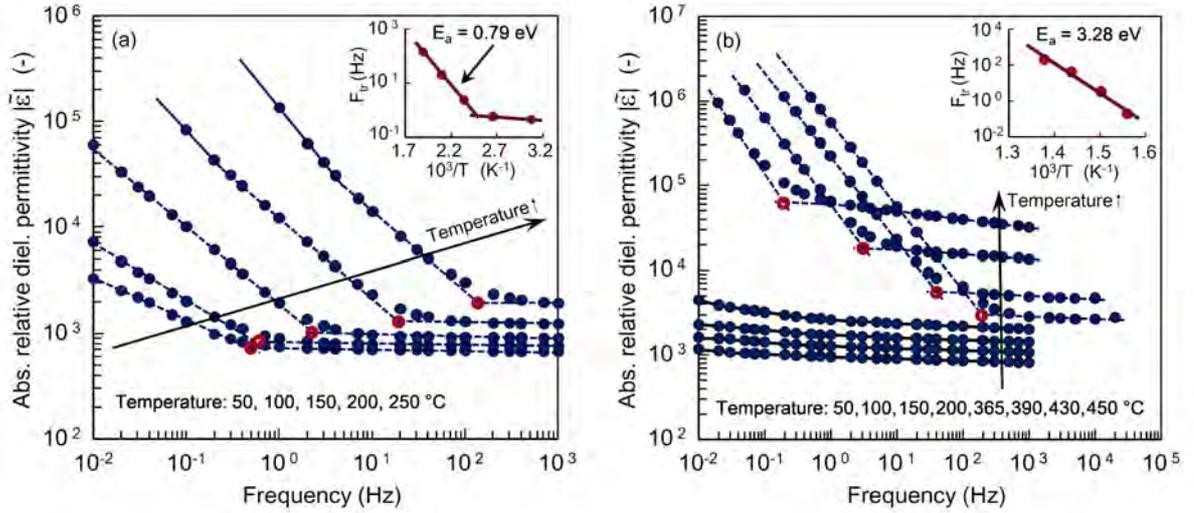

Figure 3 (color online). The frequency dispersion of the absolute value of the complex relative permittivity for (a) undoped and (b) 1.0 at% Nb-doped soft PZT (58/42) ceramics. Insets show the temperature dependence of transition frequencies in the Arrhenius scales. The transition frequencies taken for activation energies estimation are marked with full dots at the intersection of dotted lines. In (b) only data taken at selected temperatures were shown for clarity, while the measurements were carried out also at temperatures not shown in the figure.

In contrast to undoped material, the soft PZT ceramics demonstrate relatively abrupt transition of $|\tilde{\varepsilon}|$ to the frequency dispersive regime (Fig. 3b) but this happens only at temperatures near and above the Curie temperature, where interpretation of the data becomes delicate. Though the transition between the low and high frequency relaxation demonstrates a thermally activated process, a further analysis of this data representation did not lead us to any convincing conclusions. The Arrhenius plots of the transitional frequencies are not linear (e.g., see the inset in Fig. 3b). Moreover, estimated activation energies do not follow a regular trend with respect to dopant concentration (not shown). Importantly, the low frequency dielectric loss in the temperature region from 50°C to at least 200°C is much higher in hard (Fig. 2a) than in soft samples (Fig. 3b). This is in agreement with the general observation that losses in soft materials are dominated by domain walls contributions rather then by conductivity.[1,61]

The qualitatively different dielectric behavior of hard and soft ceramics is expected due to differences in their defect chemistry. It is also in agreement with the starting hypothesis that the charge transport and ageing are closely related. Hard materials are characterized by strong ageing and, as demonstrated here, by strong charge hopping, probably of ionic type. The hoping conductivity in hard materials is significant at temperatures well below Curie temperature $T_C$ where charges can interact with polarization within domains leading to ageing. The soft materials do not exhibit significant ageing. The steep increase in $|\tilde{\varepsilon}(\omega)|$ evidenced in soft PZT at low frequencies and at temperatures close to and above $T_C$ probably has, as will be discussed later, a different origin than the one in hard materials at temperatures well below $T_C$. Ionic hopping conduction is absent (or negligible) in soft PZT below Curie temperature. This result was anticipated as there are no mobile ionic defects in soft materials; it is reasonable to assume that Pb vacancies associated with donor dopants are less mobile than oxygen vacancies associated with acceptor dopants. The weak aging in soft PZT is thus in agreement with the absence of significant contribution to the conductivity of mobile ionic charge carriers. Because the conductivity in soft materials, whatever is its origin, is significant only at temperatures close or above Curie point, it can have little influence on stabilization of domain structure.

To give further support to these conjectures we next compare activation energies for the conduction and ageing in hard samples. We first analyze the activation of the dc conduction in the soft and hard materials by plotting the real part of the ac conductivity [eq. (2)], $\sigma'(\omega) = \omega \widetilde{\varepsilon}''(\omega) = \sigma_0 + \omega \varepsilon''(\omega)$ as a function of frequency and temperature, Fig. 4. At lower frequencies $\sigma'(\omega)$ should approach the frequency independent dc conductivity, $\sigma_0$. A tendency to such behavior is observed in both soft and hard samples, in particular at high temperatures, as shown in Fig. 4. We emphasize that at lower temperatures, $\sigma'(\omega)$ continues to slowly change even at the lowest frequencies examined. There are two reasons for such behavior. One is that true $\sigma_0$ [nearly flat $\sigma'(\omega)$] is not reached but would be observed if measurements were made at even lower frequencies. This means that $\sigma_0$ determined by extrapolation from these graphs (see Figure 4) is somewhat overestimated. The other possibility was discussed in detail in Ref. 44, where it was proposed that a slowly varying $\sigma'(\omega)$ indicates a polarization mechanism related to longer-range charge migration. In either case, the slow evolution of the ac conductivity suggests charge migration over an extended range; extrapolation of $\sigma_0$ from the nonflat part of $\sigma'(\omega)$ is then just a rough estimate of the contribution of this long-range process to the apparent $\sigma_0$. For simplicity, in the rest of the text we shall not differentiate between these two possibilities and will refer to both either as dc conductivity or long-range charge migration.

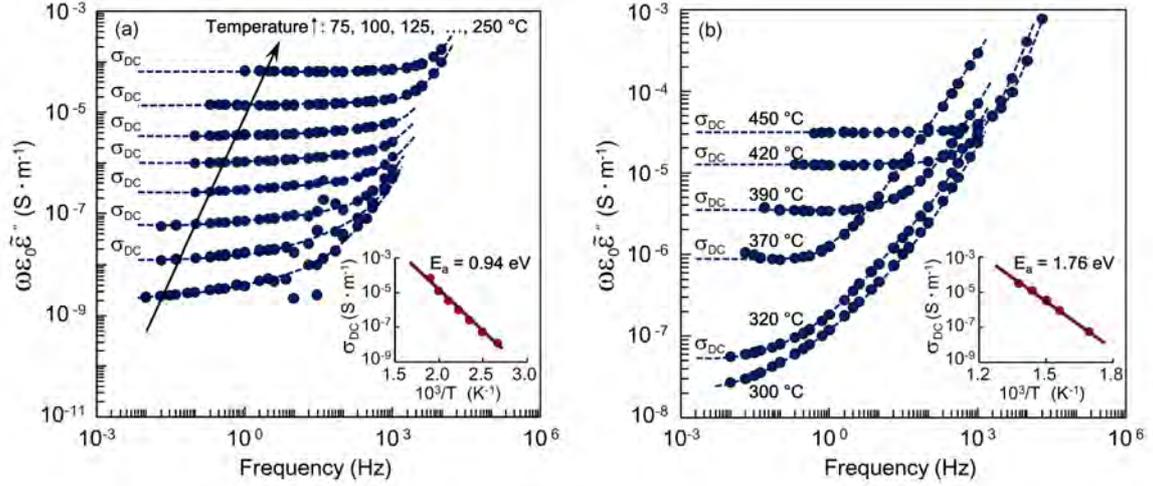

Figure 4 (color online). The frequency dispersion of the $\omega\varepsilon_0\tilde{\varepsilon}''$, which approaches the dc conductivity $\sigma_0$ at low frequencies: (a) for hard (1 at.% Fe-doped) and (b) for soft (1 at. % Nb-doped) PZT (58/42) ceramics. The insets show the temperature dependences of the extrapolated dc conductivity $\sigma_0$ in the Arrhenius scales and corresponding activation energies.

The activation energies for the dc conduction (or long range charge migration) have been determined for all investigated hard and soft PZT ceramics. Examples are shown in insets of Fig. 4. The summary of the activation energies for the ac (short range) and dc (longer-range) charge migration in hard (ac and dc conductivity) and soft PZT (dc conductivity only) is presented in Fig. 5. The activation energies for the dc conduction in soft ceramics are found to be about two times higher than in hard PZT. The value of 1.76 eV in soft PZT is identical to that reported for bulk conductivity of soft PZT above $T_C$ by Verdier et al.[49] and is equal to the half of the energy gap expected in PZT. Therefore, the dc conductivity in soft PZT is may be dominated by the intrinsic electronic conduction rather than conductivity of lead vacancies. For both hard and soft materials activation energies for the dc conduction slightly decrease with increasing dopant concentration. In contrast, the activation energy for the ac conduction in hard PZT slightly increases with increasing dopant concentration. We do not have at present an explanation for the concentration dependence of the conductivity.

In hard materials, the difference between the activation energies for the ac and dc processes [$E_a(dc) > E_a(ac)$] can be explained by the hopping nature of the conduction, the nature of charged defects responsible for the conductivity and the distance over which charge carriers move. In all hoping processes the activation energy for the dc conductivity is larger than for ac because for long range migration charges need to overcome the largest potential barriers.[45] The obvious candidate for hopping charge migration in hard materials are oxygen vacancies associated with $Fe'_{Ti} - V_O^{\bullet\bullet}$ defect pairs; as mentioned earlier in this section, hopping via $Fe^{+2}$ and $Fe^{+3}$ cations is not supported by EPR experiments. Thus, the local jumps of oxygen vacancies $V_O^{\bullet\bullet}$ around $Fe'_{Ti}$ (short-range hopping) occur under an ac field unless the bonds in the defect associates are broken and the long-range ("dc") conduction is activated as a continuous charge drift over long distances. Assuming the mobility of oxygen vacancies as the major factor influencing the conduction at lower temperatures, the difference in activation energies for the ac and dc conduction then arises from the necessity of breaking the bond of the pair for longer-range migration, in addition to overcoming other barriers encountered during the long-range migration.[45]

The analysis of ac and dc conductivity and associated activation energies in soft samples is more difficult because the measurable data for $\sigma_0$ and transition region where ac conductivity begins to dominate effective permittivity appear close to the Curie temperature. Nevertheless, based on the literature[49] and our own data (dielectric dispersion, absence of aging) one can state that the charge transport mechanism in soft materials is qualitatively different from the one in hard samples, possibly consisting of a dominant electronic conduction rather than ionic conductivity of lead vacancies.

To make a link between the aging and conductivity we now compare activation energies of conductivity for hard PZT obtained in this study with the activation energies for loop depinching of the same base compositions of PZT ceramics reported by Carl and Härdtl.[10] As shown in Fig. 5d, the activation energy (0.55 -0.7 eV) for deaging (or more precisely, for loop depinching part of the deaging process) by field cycling with switching ac field[10] is energetically comparable with the activation energy for charge hopping conduction (0.6-0.8 eV). Note that this energy probably includes energy for the movement of the vacancy within oxygen octahedra and hopping to neighboring octahedra. The activation energies assigned to oxygen vacancy hopping in this work are considerably lower than 1.2 eV calculated by Arlt and Neumann for diffusion of oxygen vacancies in PZT.[26] On the other hand they are higher than energies predicted by *ab-initio* calculations needed for displacement of an oxygen vacancy within oxygen octahedron in compositions on the tetragonal side of the PZT phase diagram.[68]

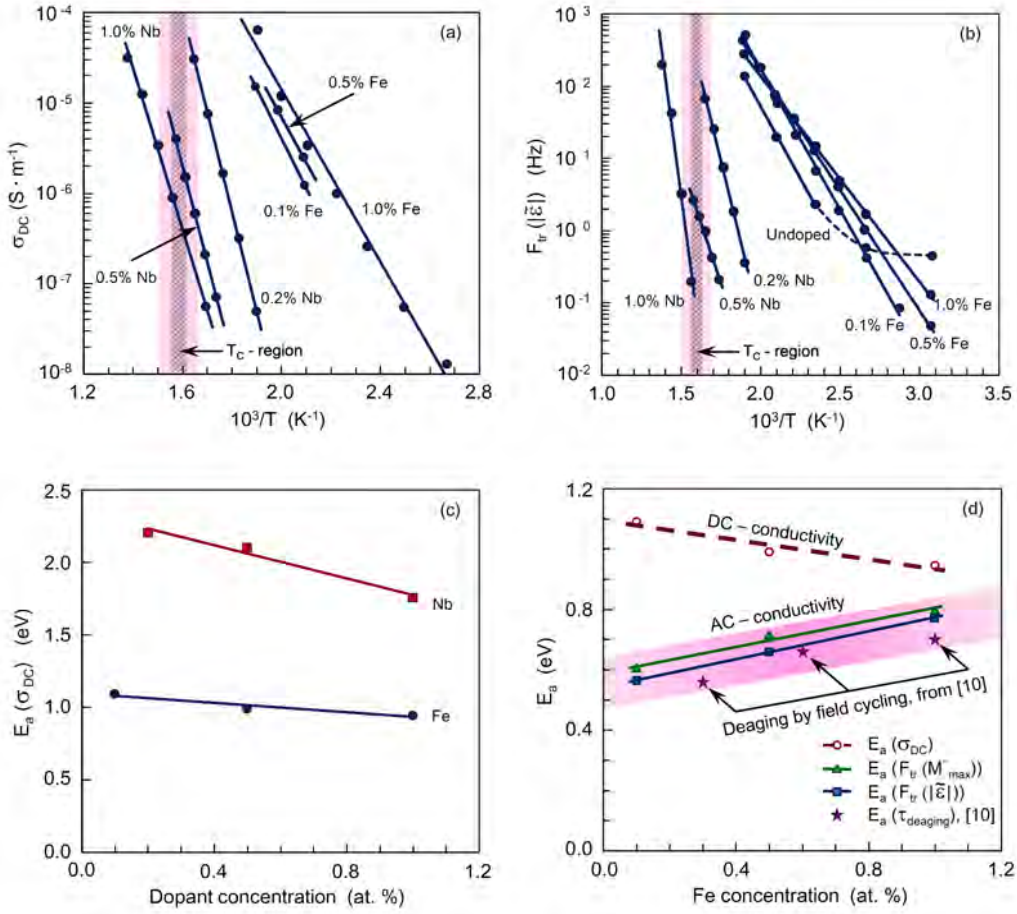

Figure 5 (color online). Summary of dielectric relaxation parameters in the hard and soft PZT 58/42 ceramics: (a) temperature dependence of the estimated dc-conductivity for different dopant concentrations; (b) transition frequencies determined from $|\tilde{\varepsilon}(\omega)|$ curves; (c) activation energies of the estimated dc-conductivity versus dopant concentration; (d) activation energy of the ac hopping conduction evaluated by different methods for various Fe concentration in hard PZT. Deaging data in (d) are taken from Ref. 10.

Our results on the ac and dc conductions in hard PZT are in agreement with the general statement about the hopping conduction in many solids: the shorter the distance within which the charges move, the smaller the activation energy for such mobility.[45] Among the three aging-deaging scenarios the shortest distance required for the electrostatic rearrangement of charged defects is assumed by the bulk effect i.e ordering - disordering of the defect dipoles with respect to the spontaneous polarization between domain walls. Thus, the comparison of the activation energies shown in Fig. 5d indicates that the deaging in hard PZT occurs at conditions sufficient for the activation of dipoles reorientation but insufficient for the activation of a long-distance charge drift process. This result leads us to the following conclusion: the aging-deageing process, which in hard PZT manifests itself in hysteresis pinching-depinching, is associated with the local movements of defect dipoles via short-range hopping of oxygen vacancies rather than with the long range charge drift mechanism. However, as shown in Ref. 37 depinched loops may be further open by large field cycling at higher temperatures. This is evidence that a process with higher activation energy such as a mechanism involving long-range charge redistribution may contribute to deaging under large fields.

## 5. Summary and further implications

The electric conductivity and hardening (which is the same as ageing in hard PZT) appear to be intimately related. Short range migration of charges (in this case oxygen vacancies) is needed to align defects dipoles with polarization within domains and thus immobilize domain walls. Unlike electric fatigue,[31,69] though, the ageing does not have to take place under electric field. The driving force for the defect dipoles reorientation is minimization of their elastic and electrostatic energies. The conductivity and aging are, however, related through the same mechanisms of charge migration: the short range hopping of oxygen vacancies with activation energies sufficient to account for many features of the aging-deaging process (e.g., loop pinching/depinching), and probably long range charge drift, which appears necessary to complete the deaging process under electric field.

While the nature of defects that immobilize domain walls in hard ceramics now seems to be well established (oxygen vacancy-cation acceptor pair), their emplacement is still being debated. It has been suggested through *ab-initio* calculations that aging and domain wall pinning have origin in oxygen vacancies that are situated within the domain wall region [28,33] and not between domain walls. If this is indeed the case, it might imply the following. A PZT sample is always prepared at high temperatures (600°C for thin films, 1200°C for ceramics). It is reasonable to assume that $Fe'_{Ti} - V_O^{\bullet\bullet}$ dipoles already exist when the sample is cooled through the Curie temperature where domain walls are formed. The hardening starts at this point. Considering the short range charge migration needed for ageing and hardening, it would be interesting to explore whether domain walls form around existing $Fe'_{Ti} - V_O^{\bullet\bullet}$ dipoles or isolated vacancies migrate to the wall region. The latter scenario leaves open the question of the driving force needed for disrupting the $Fe'_{Ti} - V_O^{\bullet\bullet}$ bond, since $Fe'_{Ti}$ are unlikely to move. From this perspective, position of dipoles between the domain walls rather than within domain walls appears to be more plausible.

In hard materials, the ageing and hardening have been correlated to the sufficiently high ionic conductivity below $T_C$. It is thus not surprising that the absence of aging in soft materials can be correlated to the relatively low conductivity below $T_C$. As shown here, the charge mobility is limited in soft PZT below $T_C$ and stabilization of the domain walls structure by the migration of charged defects is not expected. Experimental results presented for soft PZT give no evidence that defect dipoles such as $(Nb_{Ti}^{\bullet} - V_{Pb}'')'$ (if they are formed at all) align with polarization or that $V_{Pb}''$ moves by hopping. While this is in agreement with the *absence* of hardening in soft materials, it does not explain the softening (higher properties with respect to undoped material). Our experiments show that undoped PZT behaves as a weakly hard material and that even PZT doped with 0.2% Nb exhibits slight pinching of the polarization-electric field hysteresis loop.[70] This agrees with the early suggestion that addition of Nb to PZT compensates for naturally present acceptor impurities.[1] Whether donor dopants possess beyond simple charge compensation an additional softening effect through stress relief [1,40,61] (leading to very rapid ageing rather than its absence) or through another elusive process remains at present unanswered. Importantly, one should not expect that fully or over-compensated donor doped PZT exhibits properties that would be observed in defect-free PZT because the domain wall structure of two materials would be different.

**Acknowledgements:**

Authors acknowledge Prof. H. Tuller for a useful discussion regarding the present work and Li Jin for comments on the manuscript. This work was financially supported by the Swiss National Science Foundation Projects 200021-116038 and 200020-124498.

**References:**

[1] B. Jaffe, W. R. Cook, and H. Jaffe, *Piezoelectric Ceramics* (Academic, New York, 1971).
[2] L. E. Cross "Ferroelectric Ceramics: Tayloring Properties for Specific Applications", in *Ferroelectric Ceramics*, edited by N. Setter and E. L. Colla (Birkhäuser, Basel, 1993),
[3] W. A. Schulze and K. Ohino, Ferroelectrics **87,** 361 (1988).
[4] L. X. Zhang and X. Ren, Phys. Rev. B **71,** Art. No. 174108 (2005).
[5] P. V. Lambeck and G. H. Jonker, Journal of Physics and Chemistry of Solids **47,** 453 (1986).
[6] Y. A. Genenko, Physical Review B (Condensed Matter and Materials Physics) **78,** Art. No. 214103 (2008).
[7] Y. A. Genenko and D. Lupascu, Phys. Rev. B **75,** Art. No. 184107 (2007).
[8] D. C. Lupascu, Y. A. Genenko, and N. Balke, J. Am. Ceram. Soc. **89,** 224 (2006).
[9] M. E. Drougard and D. R. Young, Phys. Rev. **94,** 1561 (1954).
[10] K. Carl and K. H. Haerdtl, Ferroelectrics **17,** 473 (1978).
[11] F. A. Kroger and H. J. Vink, Solid State Physics-Advances in Research and Applications **3,** 307 (1956).
[12] L. X. Zhang and X. B. Ren, Physical Review B **73,** Art. No. 094121 (2006).
[13] Z. Feng and X. Ren, PHYSICAL REVIEW B **77,** Art. No. 134115 (2008).
[14] A. Misarova, Sovet Phys. - Solid State **2,** 1160 (1960).
[15] P. V. Lambeck and G. H. Jonker, Ferroelectrics **22,** 729 (1978).
[16] E. T. Keve, K. L. Bye, A. D. Annis, and P. W. Whipps, Ferroelectrics **3,** 39 (1960).
[17] K. Okada, J. Phys. Soc. Japan **16,** 414 (1961).
[18] M. E. Lines and A. M. Glass, *Principles and Applications of Ferroelectrics and Related Materials* (Clarendon, Oxford, 1979).
[19] U. Robels and G. Arlt, J. Appl. Phys. **73,** 3454 (1993).

20   W. L. Warren, K. Vanheusden, D. Dimos, G. E. Pike, and B. A. Tuttle, J. Am. Ceram. Soc. **79,** 536 (1996).
21   W. L. Warren, G. E. Pike, K. Vanheusden, D. Dimos, B. A. Tuttle, and J. Robertson, J. Appl. Phys. **79,** 9250 (1996).
22   W. L. Warren, D. Dimos, G. E. Pike, K. Vanheusden, and R. Ramesh, Appl. Phys. Lett. **67,** 1689 (1995).
23   L. X. Zhang, E. Erdem, X. B. Ren, and R. A. Eichel, Applied Physics Letters **93,** Art. No. 202901 (2008).
24   A. S. Nowick and W. R. Heller, Adv. Phys. **14,** 101 (1965).
25   U. Robels, C. Zadon, and G. Arlt, Ferroelectrics **133,** 163 (1992).
26   G. Arlt and H. Neumann, Ferroelectrics **87,** 109 (1988).
27   X. B. Ren, Nature Materials **3,** 91 (2004).
28   C. H. Park and D. J. Chadi, Phys. Rev. B **57,** R13961 (1998).
29   V. S. Postnikov, V. S. Pavlov, and S. K. Turkov, J. Phys. Chem. Sol. **31,** 1785 (1970).
30   V. S. Postnikov, V. S. Pavlov, and S. K. Turkov, Soviet Phys. - Solid State **10,** 1267 (1968).
31   W. L. Warren, D. Dimos, B. A. Tuttle, R. D. Nasby, and G. E. Pike, Appl. Phys. Lett. **65,** 1018 (1994).
32   W. L. Warren, D. Dimos, and B. A. Tuttle, J. Am. Ceram. Soc. **77,** 2753 (1994).
33   L. He and D. Vanderbilt, Phys. Rev. B **68,** Art. No. 134103 (2003).
34   M. Takahashi, Jpn. J. Appl. Phys. **9,** 1236 (1970).
35   E. Li, H. Kakemoto, T. Hoshina, and T. Tsurumi, Jpn. J. Appl. Phys. **47,** 7702 (2008).
36   M. Morozov, PhD Thesis, Swiss Federal Institute of Technology - EPFL, 2005.
(http://library.epfl.ch/theses/?nr=3368)
37   M. I. Morozov and D. Damjanovic, Journal of Applied Physics **104,** Art. No. 034107 (2008).
38   M. Morozov, D. Damjanovic, and N. Setter, J. Europ. Ceram. Soc. **25,** 2483-86 (2005).
39   Q. Tan, J.-F. Li, and D. Viehland, Phil. Mag. B **76,** 59 (1997).
40   R. Gerson, Journal of Applied Physics **31,** 188 (1960).
41   L. Eyraud, B. Guiffard, L. Lebrun, and D. Guyomar, Ferroelectrics **330,** 51 (2006).
42   H. Mestric, R. A. Eichel, T. Kloss, K. P. Dinse, S. Laubach, P. C. Schmidt, K. A. Schonau, M. Knapp, and H. Ehrenberg, Physical Review B **71,** Art. No. 134109 (2005).
43   *Impedance Spectroscopy Theory, Experiment, and Applications*; edited by E. Barsoukov and J. R. Macdonald (Wiley, Hoboken, NJ, 2005).
44   A. K. Jonscher, *Dielectric relaxation in solids* (Chelsea Dielectric Press, London, 1983).
45   J. C. Dyre, Journal of Applied Physics **64,** 2456 (1988).
46   A. R. Von Hippel, *Dielectrics and Waves* (MIT Press, Cambridge, MA., 1954).
47   A. Seal, S. Das, R. Mazumder, and A. Sen, Journal of Physics D-Applied Physics **40,** 7560 (2007).
48   A. Khodorov, S. A. S. Rodrigues, M. Pereira, and M. J. M. Gomes, Journal of Applied Physics **102,** Art. No. 114109 (2007).
49   C. Verdier, F. D. Morrison, D. C. Lupascu, and J. F. Scott, Journal of Applied Physics **97** Art. No. 024107 (2005).
50   D. L. Sidebottom, B. Roling, and K. Funke, Physical Review B **63,** Art. No. 024301 (2000).
51   V. Porokhonskyy, L. Jin, and D. Damjanovic, Applied Physics Letters **94,** Art. No. 212906 (2009).
52   D. Damjanovic, Phys. Rev. B **55,** R649 (1997).
53   D. V. Taylor and D. Damjanovic, J. Appl. Phys. **82,** 1973 (1997).
54   E. Pérez-Enciso, N. Agraït, and S. Vieira, Phys. Rev. B **56,** R2900 (1997).
55   P. Mokry, Y. Wang, A. K. Tagantsev, D. Damjanovic, I. Stolichnov, and N. Setter, Physical Review B (Condensed Matter and Materials Physics) **79,** Art. No. 054104 (2009).
56   N. A. Pertsev and G. Arlt, Journal of Applied Physics **74,** 4105 (1993).
57   U. Bottger and G. Arlt, Ferroelectrics **127,** 95 (1992).
58   G. Arlt, U. Böttger, and S. Witte, Ann. Phys. **3,** 578 (1994).
59   R. A. Eichel, H. Mestric, K. P. Dinse, A. Ozarowski, J. van Tol, L. C. Brunel, H. Kungl, and M. J. Hoffmann, Magnetic Resonance in Chemistry **43,** S166 (2005).
60   R. A. Eichel, Journal of the American Ceramic Society **91,** 691 (2008).
61   R. Gerson, Journal of Applied Physics **31,** 1615 (1960).
62   Y. Gao, K. Uchino, and D. Viehland, Jpn. J. Appl. Phys. **45,** 9119 (2006).
63   W. Kleemann, Annual Review of Materials Reserach **37,** 415 (2007).
64   I.D. Raistrick *"The electrical analogs of physical and chemical processes"* in *Impedance Spectroscopy : Emphasizing Solid Materials and Systems*; edited by J. R. MacDonald (Wiley, New York, 1987).
65   R. Richert, J. Non-Crystalline Solids **305,** 29 (2002).
66   P. B. Macedo, C. T. Moynihan, and R. Bose, Phys. Chem. Glasses **13,** 171 (1972).
67   S. R. Elliott, J. Non.Cryst. Solids **170,** 97 (1994).
68   R. A. Eichel, private communication
69   W. L. Warren, B. A. Tuttle, and D. Dimos, Appl. Phys. Lett. **67,** 1426 (1995).
70   D. Damjanovic "Hysteresis in piezoelectric and ferroelectric materials", Chapter 3 in *Science of Hysteresis*, *Vol*. *III*, edited by G. Bertotti and I. Mayergoyz (Elsevier, Amsterdam, 2005), p. 337-465.